\documentclass[superscriptaddress,prb,twocolumn]{revtex4}

\usepackage{graphicx}% Include figure files
\usepackage{verbatim}
\usepackage{amsmath,amssymb,relsize}

\def\ket#1{\mathinner{|{#1}\rangle}}

\makeatletter
\def\Ddots{\mathinner{\mkern1mu\raise\p@
\vbox{\kern7\p@\hbox{.}}\mkern2mu
\raise4\p@\hbox{.}\mkern2mu\raise7\p@\hbox{.}\mkern1mu}}
\makeatother
\usepackage{amsmath}
\usepackage{amsthm}
\usepackage{amssymb}
\usepackage{rotating}
\usepackage{epstopdf}
\usepackage{bbold}
\usepackage{times}
\usepackage[none]{hyphenat}
\usepackage{color}
\usepackage{mathrsfs}

\usepackage[normalem]{ulem}

\usepackage[percent]{overpic}

\usepackage{siunitx}
\usepackage{tikz}

\usepackage{color}
\usepackage{mathrsfs}

\newcommand\restr[2]{{% we make the whole thing an ordinary symbol
  \left.\kern-\nulldelimiterspace % automatically resize the bar with \right
  #1 % the function
  \right|_{#2} % this is the delimiter
  }}

\usepackage[belowskip=-12pt,aboveskip=10pt]{caption}

\usepackage{tikz}
\usepgflibrary{arrows}

\expandafter\def\expandafter\normalsize\expandafter{%
    \normalsize
    \setlength\abovedisplayskip{10pt}
    \setlength\belowdisplayskip{10pt}
    \setlength\abovedisplayshortskip{10pt}
    \setlength\belowdisplayshortskip{10pt}
}

\begin{document}

\author{G. Pica}
\affiliation{SUPA, School of Physics and Astronomy, University of St Andrews, KY16 9SS, United Kingdom}
\author{B. W. Lovett}
\affiliation{SUPA, School of Physics and Astronomy, University of St Andrews, KY16 9SS, United Kingdom}
\author{R. N. Bhatt}
\affiliation{Dept. of Electrical Engineering, Princeton University, Princeton, New Jersey 08544, USA}
\author{T. Schenkel}
\affiliation{Ion Beam Technology Group, Lawrence Berkeley National Laboratory, Berkeley, CA 94720, USA}
\author{S. A. Lyon}
\affiliation{Dept. of Electrical Engineering, Princeton University, Princeton, New Jersey 08544, USA}

\title{Surface code architecture donors and dots in silicon with imprecise and non-uniform qubit couplings}

\begin{abstract}
A scaled quantum computer with donor spins in silicon would benefit from a viable semiconductor framework and a strong inherent decoupling of the qubits from the noisy environment. Coupling neighbouring spins via the natural exchange interaction according to current designs requires gate control structures with extremely small length scales. We present a silicon architecture where bismuth donors with long coherence times are coupled to electrons that can shuttle between adjacent quantum dots, thus relaxing the pitch requirements and allowing space between donors for classical control devices. An adiabatic SWAP operation within each donor/dot pair solves the scalability issues intrinsic to exchange-based two-qubit gates, as it does not rely on sub-nanometer precision in donor placement and is robust against noise in the control fields. We use this SWAP together with well established global microwave Rabi pulses and parallel electron shuttling to construct a surface code that needs minimal, feasible local control.
\end{abstract}
\maketitle 
\section{Introduction}
In 1998 Loss and DiVincenzo~\cite{PhysRevA.57.120} proposed a scheme for universal quantum computing with electron spins in semiconductor quantum dots, and Kane~\cite{kane98} presented an alternative scheme with donor spin qubits. These blueprints have inspired a great deal of progress in controlling bulk donor spins~\cite{tyryshkin12,clocktransitions} and single- or few-spin donor devices~\cite{morello10,Pla2012,esrlong,buch13}, as well as dot-based quantum devices~\cite{Maune2012,PhysRevLett.108.046808,veldhorst,veldhorst2015}, but their full implementation still faces significant fundamental hurdles. In a large quantum register the delicate conditional-phase gate built between two neighbouring spins by their natural exchange interaction requires tight inter-qubit distances~\cite{kane98}, high precision local tuning of each coupling, and extreme robustness to noise~\cite{electricnoise}.

Here we show in detail how bismuth donors in silicon can be combined with quantum dots to implement a scaled surface code processor architecture that implements effective error correction~\cite{PhysRevA.86.032324} and can tolerate inaccuracy in local spin placement and tuning. The more relaxed length scales typical of quantum dots and the possibility of moving electrons short distances between them~\cite{Maune2012,PhysRevLett.108.046808,veldhorst2015,PhysRevLett.94.126802,movingdot} allow neighbouring donor spins to be implanted at least $\sim\SI{1}{\micro\meter}$ apart while retaining all their specific advantages~\cite{saeedi13,spinmem,proposal}: the pitch requirements are thus much more attainable than the $\sim\SI{20}{\nano\meter}$ donor distance required by Kane's scheme, and allow space for enough gates to adequately control wave functions in quantum dots. Scalability is further improved by using the exchange interaction between the electron spins of a donor/dot pair only to swap quantum states through adiabatic transfer, with no need to build the fragile two-qubit dynamical phases common to both of the previous proposals. This adiabatic SWAP is shown to be insensitive to large variations in donor to dot coupling, and provides the building block for the surface code CNOT that is used for error diagnosis: by construction, this pivotal operation does not need individual voltage tuning or precise timing, and is very robust against electric and magnetic noise. Our proposal is completed by microwave-driven spin rotations of the Bi donors, whose fidelities have recently surpassed the fault-tolerance threshold \cite{donorfidelity}. These fields are applied globally, while the only site-selectivity required is provided by feasible local electric control: thus parallel processing of a large number of two-qubit gates in silicon becomes more viable.

We improve donor-only architectures with respect to i) the enlarged distance between donors, that could even be increased to several microns if more area is required for readout and classical control circuitry, with little increase in decoherence; ii) the robustness of the CNOT gate to the strong variations in the magnitude of the exchange coupling as the donor separation changes~\cite{oscilla2}. At the same time, a dot-only architecture would miss the crucial benefits inherent to the nuclear spin, which we show allows for selective donor/dot entanglement within the SWAP that we envision, thanks to the electron-nuclear spin hybridization achieved by the natural hyperfine interaction at the donor. The possibility of coupling Zeeman-split dot and coherent donor states is exclusively provided by Bismuth atoms that have a large enough nuclear spin space (nuclear spin $I$= 9/2). 
\section{The architecture and the protocol}
\begin{figure}[t!]
     \includegraphics[width=.48\textwidth]{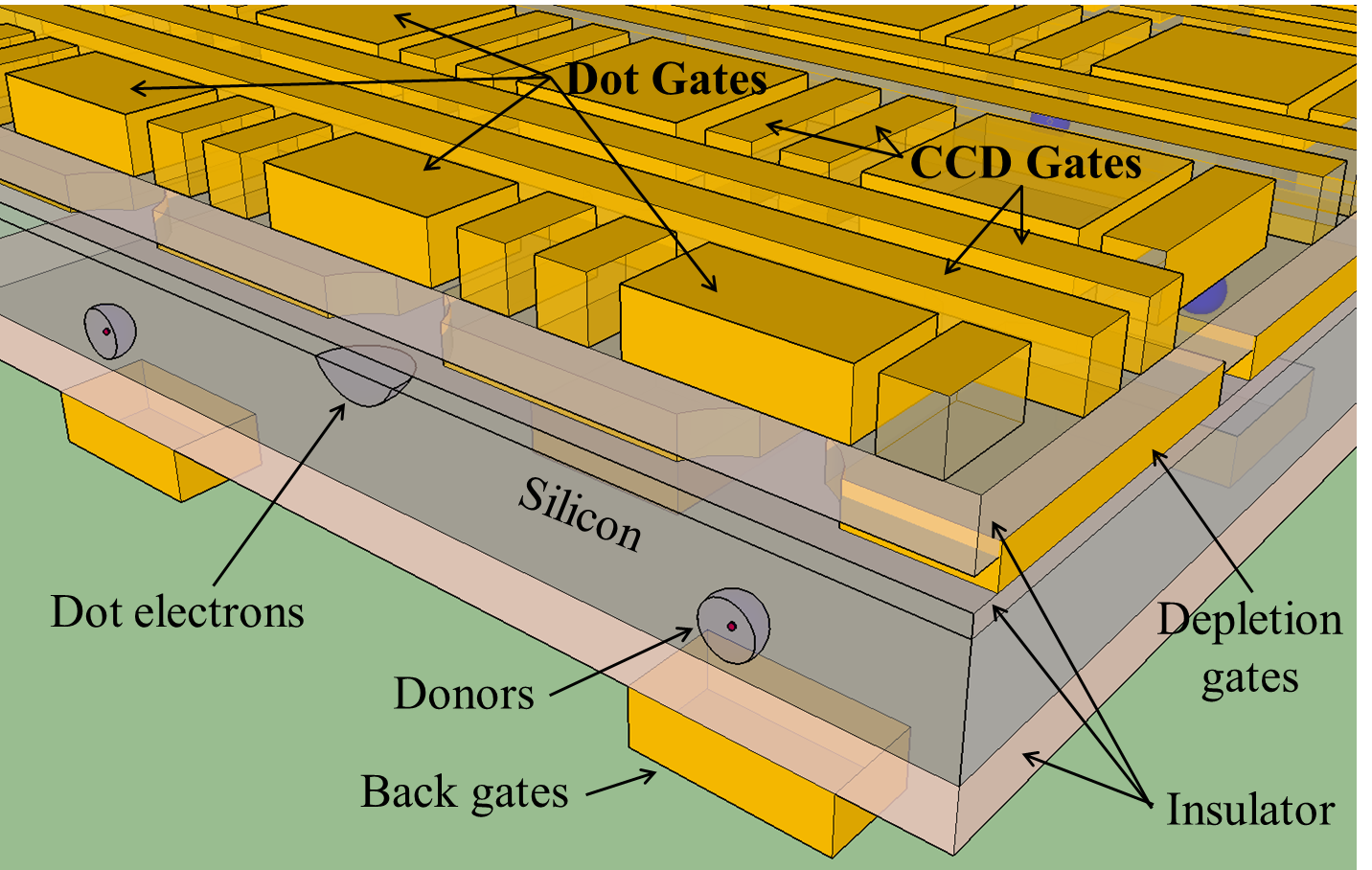}
 \caption{Schematic diagram of the donor-dot array structure. The combination of top gates and holes in the depletion gate form the quantum dots, half of which are occupied with data qubit electrons.  These electrons can be moved to dots positioned above the donor measurement qubits, each with a back gate to control the exchange coupling between the donor electron and the electron in the quantum dot.}
  \label{fig1}
\end{figure}

In Fig.~\ref{fig1} we show an idealized diagram of a portion of the structure we are suggesting.  Here we assume that the donors are incorporated below the quantum dots, and the donor/dot interaction is controlled by a back gate, as in the devices suggested by Schenkel \emph{et al.}~\cite{Schenkel2013}. This vertical donor-dot configuration allows the dots to have quite simple gate structures and easily modeled electrostatic fields. All electrodes in the upper layer of Fig.~\ref{fig1} act as 3-phase charge-coupled device (CCD) gates~\cite{ccd}, moving all dot electrons in unison -- this requires only five independent gate signals: two for horizontal shuttling, two for vertical coupling, and one for dot confinement.  The underlying back gates are individually addressable to determine which qubits are involved in each surface code cycle. Crucially, the robustness of the adiabatic SWAP to donor-dot coupling strength variations implies that the back gates can all be switched between two standard voltages, rather than requiring individual tuning at every site. Other layouts have technological advantages and disadvantages which are discussed in Appendix D.

%\begin{figure}[h!]
  %\centering
    %\includegraphics[width=.45\textwidth]{fig2}
  %\caption{Spin transition energies of neutral Si:Bi donors (color curves) and an electron in a quantum dot in Si (black straight line) as a function of magnetic field. The dot and donor transition energies are nearly degenerate at the 5 GHz clock transition of the donor. Every donor curve represents two transitions, one `allowed' and one `forbidden' in the high field limit~\cite{natureactivation}, whose separation is not resolved in the picture. The inset shows a calculation of the mixing between the `allowed' and quantum dot transitions when the donor and dot are exchange coupled.}
  %\label{fig2}
%\end{figure}

The surface code architecture we will consider consists of a square planar array of qubits as described in detail by Fowler \emph{et al.}~\cite{PhysRevA.86.032324}. We consider the data qubits (DQ) to be the spin of the electrons in the quantum dots, and the measurement qubits (MQ) to be states of the donor electron and nuclear spins (coupled through the hyperfine interaction). There are four basic operations which the qubit array must perform for error correction: (1) movement of the entire array of DQ to each of the four adjacent MQ in turn, (2) addressable CNOT operations with the DQ as control and the MQ as targets, (3) measuring the MQ, and (4) applying global Hadamard gates to the DQ. This protocol allows the diagnosis of phase-flip and bit-flip errors accumulating in the DQ array. The movement operation (1) would utilize the surface gates described above. When electrons are released from their quantum dots and allowed to move in a two-dimensional layer, it is known that their spin coherence is significantly degraded, but still of the order of microseconds~\cite{PhysRevLett.94.126802}. If the period of the donor-dot array is of order a micron, and the electrons are moved at $10^{6}$ cm/s (about 500 mK electron energy in Si), the time to transport the electrons is only about \SI{100}{\pico\second}. The error per qubit accumulated during the electron motion is $<10^{-4}$, less than that which can be expected from gate operations. For operation (3) we assume the ability to measure the spin state of individual donors. Accurate measurement of electron and nuclear spins of single donors in Si has recently been demonstrated~\cite{morello10,Pla2013} using spin-to-charge conversion via electron tunnelling, though selective optical excitation with relaxed donor positioning constraints may also be possible~\cite{cheuknew}. The tunnelling measurements were performed by coupling a single-electron-transistor (SET) to a donor implanted within a $90\times90$ \SI{}{\nano\meter}$^{2}$ region~\cite{morello10,Pla2013}. For this style of donor readout the SET would be integrated with the bottom gate and the size of the readout structures would be well within the micron pitch of our proposed architecture. The Hadamard operation (4) will use a global microwave field. Such operations can be performed with a fidelity better than $99.6\%$~\cite{veldhorst}.

The main obstacle to the implementation of a surface code is thus performing the addressable CNOT gate between MQ and DQ, which we now construct using adiabatic transfer swaps based on the exchange coupling between the donor and its paired dot electron (see Fig.~\ref{fig1}). We start with a detailed description of the Hamiltonian of the pair and focus on the selection rules induced by the hyperfine-coupled nuclear spin on the exchange-coupled electron states of the donor and the dot.

\section{The logical Hilbert space} The Hilbert space spanned by the tensor combinations of the states of the donor nuclear spin $I=9/2$, the donor electron spin $S_{\text{donor}}=1/2$ and the dot electron spin $S_{\text{dot}}=1/2$ is 40-dimensional, and is governed by the Hamiltonian 
\begin{equation}\label{levelstot}
H=\gamma_{e} \textbf{B}_{0}\cdot\textbf{S}_{\text{donor}}-\gamma_{n} \textbf{B}_{0}\cdot\textbf{I}+A \hspace{.5mm} \textbf{S}\cdot\textbf{I}+\gamma_{e} \textbf{B}_{0}\cdot\textbf{S}_{\text{dot}}+J \textbf{S}_{\text{donor}}\cdot\textbf{S}_{\text{dot}},
\end{equation}
where $\textbf{B}_{0}$ is the applied dc magnetic field, $A$ is the hyperfine interaction between the nuclear spin $\textbf{I}$ and the electron spin $\textbf{S}$, $\gamma_{e}=\dfrac{g_{e} \mu_{B}}{\hbar}=\SI{27.997}{\giga\hertz/\tesla}$ is the magnetic moment of the electron, $\gamma_{n}=\SI{0.007}{\giga\hertz/\tesla}$ is the nuclear magnetic moment, and $J$ is the exchange coupling between the two electron spins. This can be increased from zero -- when the electron wavefunctions do not overlap much -- by locally tuning the back gate voltage as described above, causing the dot electron density to be pulled towards the implanted impurity.

With $J=0$, the transition energies of the uncoupled donor and dot spin states as a function of applied magnetic field are shown in Fig.~2(a). We will utilize the donor `clock transitions' whose frequency is independent of magnetic field to first order~\cite{clocktransitions}, and thus can cross the Zeeman quantum dot transition (linear in magnetic field). Fig.~2(a) shows that the dot transition crosses the lowest donor clock transition almost exactly at its minimum, where the Bi spins have particularly long coherence. This feature is unique to Bismuth among the group V donors, since as we show in Appendix A it needs the nuclear spin to be at least $9/2$.

Electron spin resonance (ESR) measurements of the clock transition near 7 GHz have shown that there are two nearly degenerate components at every clock transition~\cite{natureactivation,clocktransitions}. Since they involve different initial and final states, these four states can be used as two independent qubits residing on the bismuth. It is convenient to label the upper of each pair of transitions `forbidden' and the lower `allowed', even though this terminology is properly descriptive only in the high field limit (where Zeeman splitting is much larger than hyperfine energy).

We thus focus in our scheme on a combination of three qubits, which are shown schematically in Fig.~2(b). The first (left) is the electron spin in the dot, and given the name `Dot' qubit. With the magnetic field held near the 5 GHz clock transition ($B_{0}\approx \SI{.185}{\tesla}$), the dot qubit can be driven by conventional microwave ESR fields. The second qubit consists of the donor states at the 5 GHz clock transition. In the high-field limit this qubit would be simply an electron spin on the donor, and thus we call it the `ESR' qubit. The third qubit is the coupled electron and nuclear states making up the allowed versus the forbidden transitions.  In the high field limit this would just be a nuclear spin, and we call it the `NMR' qubit. The transition energy of this qubit is about 0.74 GHz.  All three qubits can be driven with microwave fields. 
\begin{figure}[t!]
  \centering
\begin{minipage}{.49\textwidth}
\begin{overpic}[width=.99\textwidth,height=7.5cm]{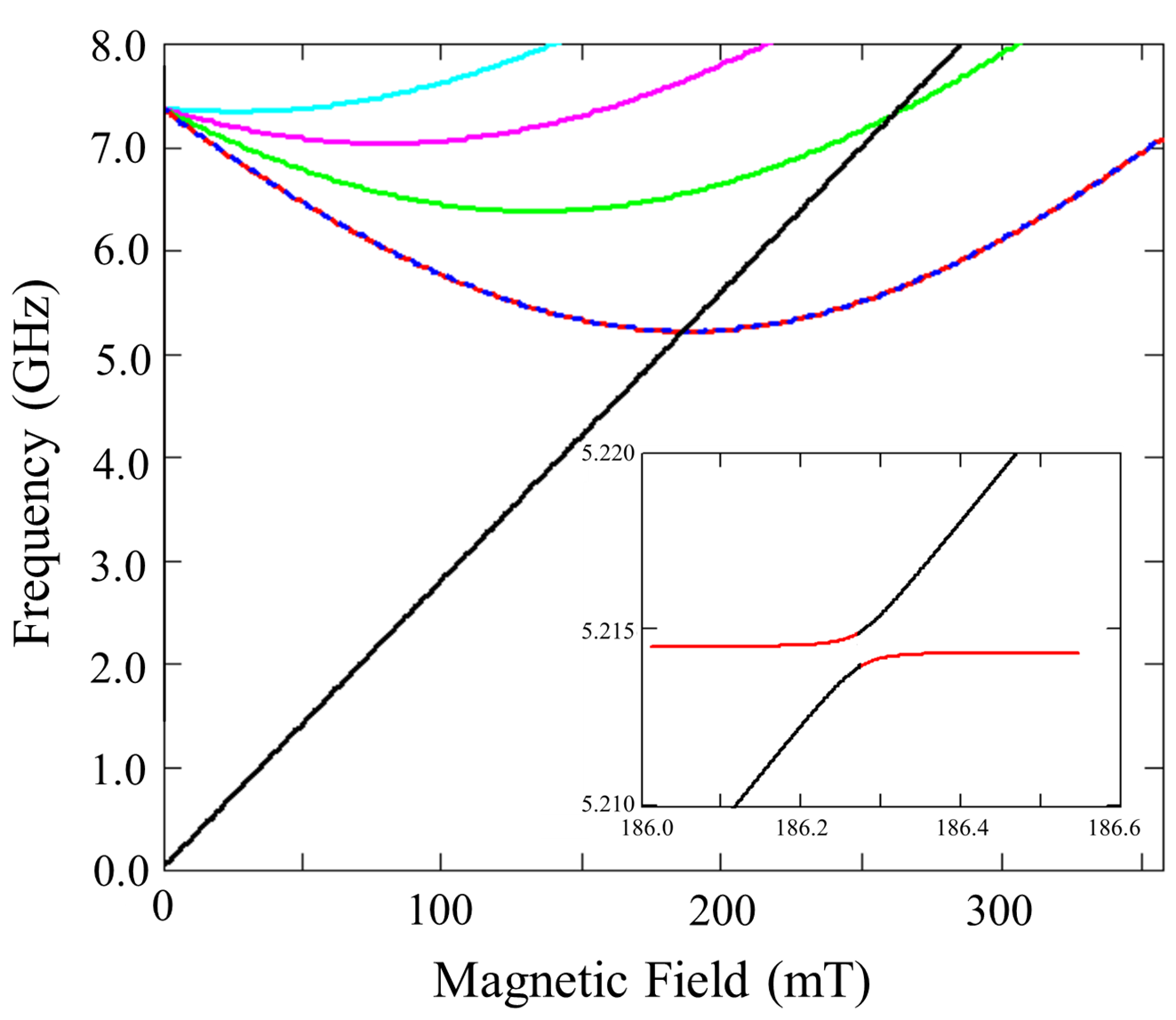}
  \put (20,45) {\Large{$(a)$}}
\end{overpic}
\end{minipage}
\begin{minipage}{.49\textwidth}
\begin{overpic}[width=.99\textwidth]{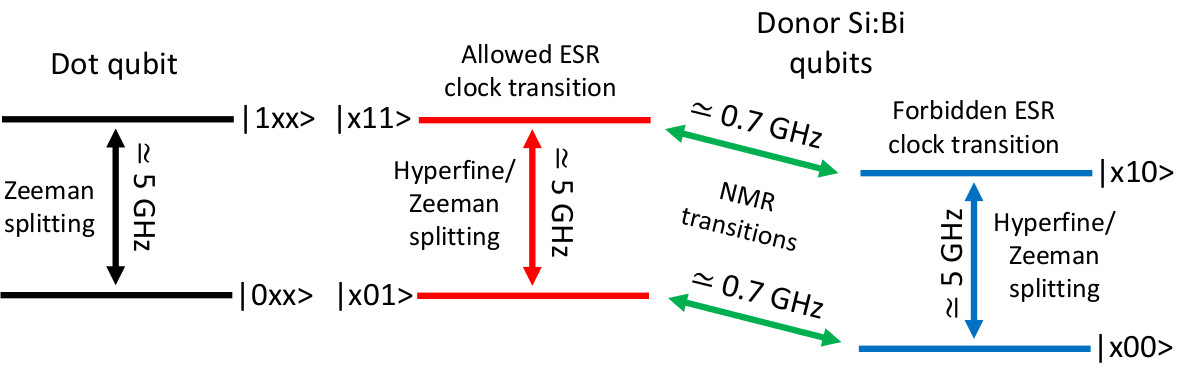}
\put (20,28) {\Large{$(b)$}}
\end{overpic}
\end{minipage}
  \caption{$(a)$ Spin transition energies of neutral Si:Bi donors (color curves) and an electron in a quantum dot in Si (black straight line) as a function of magnetic field. Every donor curve represents two transitions, one `allowed' and one `forbidden' in the high field limit~\cite{natureactivation}, whose separation is not resolved in the figure. The dot and donor transition energies cross at the 5 GHz clock transition of the donor. The inset shows a calculation of the mixing between the `allowed' and quantum dot transitions when the donor and dot are exchange coupled. $(b)$ Diagram of the transition energies of the three qubits in a donor-dot structure. The dot qubit and the allowed transition of the ESR qubit ($\ket{\text{NMR}} = \ket{1}$) can be coupled by a gate-controlled exchange interaction.}
  \label{fig2}
\end{figure}

The Hamiltonian in Eq.~\ref{levelstot} with $J=0$ can be diagonalized with separable combinations of the following dot and donor states: $ \{\ket{\downarrow},\ket{\uparrow}\}_{\text{dot}} \otimes |\pm, m\rangle_{\text{donor}}$, where the donor eigenstates are written, using the notation $\ket{S^{z}_{\text{donor}},I^{z}}$, as~\cite{PhysRevLett.105.067602} (see Appendix A for more details)
\begin{equation}\label{donoreigen}
|\pm, m\rangle = a^{\pm}_{m}|\pm 1/2, m\mp 1/2\rangle+b^{\pm}_{m}|\mp 1/2, m\pm 1/2\rangle,
\end{equation}
with $m=S^{z}_{\text{donor}}+I^{z}$ being the sum of the electron and nuclear spin projections on the quantization axis, 
\begin{eqnarray}
a^{\pm}_{m}=  \begin{cases}
                                                       \cos(\theta_{m}/2) \\
                                                             -\sin(\theta_{m}/2) 
                                  \end{cases}  ,&
b^{\pm}_{m}= \begin{cases}
                                                             \sin(\theta_{m}/2) \\
                                                             \cos(\theta_{m}/2) 
                                  \end{cases} ,
\end{eqnarray}
and
\begin{equation}
\theta_{m}=\arctan\left[{\frac{A \sqrt{I(I+1)+\frac{1}{4}-m^{2}}}{(A m+B_{0} \gamma_{e}+B_{0} \gamma_{n})}}\right], \hspace{2mm} 0\leq\theta_{m}<\pi.
\end{equation}  
The `allowed' (A) pair and the `forbidden' (F) pair of donor/dot states that cross close to the sweet spot considered here can be written, in the basis defined above, as 
\begin{equation}\label{states}\begin{array}{l}
|1\rangle^{A} = \ket{\downarrow}_{\text{dot}} \otimes \ket{+,-3}_{\text{donor}} , \\
|2\rangle^{A}= \ket{\uparrow}_{\text{dot}} \otimes \ket{-,-4}_{\text{donor}},
\\
|1 \rangle^{F} = \ket{\uparrow}_{\text{dot}} \otimes \ket{-,-3}_{\text{donor}}, \\
|2 \rangle^{F}= \ket{\downarrow}_{\text{dot}}\otimes \ket{+,-4}_{\text{donor}}.
\end{array}
\end{equation}
that we enlarge to our computational basis (with subscripts omitted for simplicity from now on)
\begin{equation}\label{statesthree}\begin{array}{ll}
\ket{1}\equiv\ket{000}=\ket{\downarrow}\ket{-, -3}& \ket{5} \equiv\ket{100}=\ket{\uparrow}\ket{-, -3}\\
\ket{2}\equiv\ket{001}=\ket{\downarrow}\ket{-, -4} & \ket{6}\equiv\ket{101}=\ket{\uparrow}\ket{-, -4} \\ \ket{3}\equiv\ket{010}=\ket{\downarrow}\ket{+, -4}& \ket{7}\equiv\ket{110}=\ket{\uparrow}\ket{+, -4}\\
\ket{4}\equiv\ket{011}=\ket{\downarrow}\ket{+, -3}& \ket{8}\equiv\ket{111}=\ket{\uparrow}\ket{+, -3}.\\
\end{array}
\end{equation}

When a nonzero exchange coupling between the dot electron and the donor electron is turned on, the crossing between $\ket{1}^{A}$ and $\ket{2}^{A}$ is avoided, as shown in the inset of Fig.~2(a), while there is no avoided crossing between the $\ket{1}^{F}$ and $\ket{2}^{F}$ states. Among all the states in Eq.~\ref{states}, the exchange interaction couples electron spin states with the same $S^{z}_{\text{donor}}+S^{z}_{\text{dot}}$ projection, but the nuclear spin further selects only the states with the same nuclear spin projection to be coupled. Thus, as it is evident from combining Eqs.~\ref{states} and~\ref{donoreigen}, ${}^{A}\langle 1| J \textbf{S}_{\text{donor}}\cdot\textbf{S}_{\text{dot}}| 2 \rangle^{A} \neq 0$, while ${}^{F}\langle 1| J \textbf{S}_{\text{donor}}\cdot\textbf{S}_{\text{dot}}| 2 \rangle^{F} = 0$. 

As a consequence, if we sweep the magnetic field through the donor/dot degeneracy point in Fig.~2(a) and at the same time pulse the donor/dot exchange coupling, the nuclear state fundamentally determines the occurrence of a population transfer between $\{\ket{1}^{A}, \ket{2}^{A}\}$ and nothing but phase accumulation between $\{\ket{1}^{F}, \ket{2}^{F}\}$. Within our computational basis this represents a natural three-qubit NMR-controlled SWAP operation (Fredkin gate~\cite{raey}), where states $\ket{4}$ and $\ket{6}$ (where the NMR qubit is in state `1' ) are SWAPped but states $\ket{3}$ and $\ket{5}$ (where the NMR qubit is in state `0' ) are not. 

In the next section we will describe how this logical gate can be combined with well established Rabi techniques to realize the surface code CNOT, and describe the detailed operations that complete a surface code cycle.

\section{Surface code cycle} At the beginning of each cycle, we assume that the dots contain the DQ and the $\ket{\text{ESR}}$ qubits have been initialized in a $\ket{0}$ state, as e.g. described in Ref.~\onlinecite{natureactivation}. The transfer gates bring the appropriate data electron to the quantum dot situated above the donor. The magnetic field will be held below that corresponding to the degeneracy of the dot and donor transitions described above [Fig.~2(a)]. Pulsing the back gate voltage below the donors selected for an operation turns on an exchange coupling between the donor and the dot. As shown in Fig.~3(a) and anticipated above, the magnetic field is swept through the resulting avoided crossing to swap the $\ket{101}$ and $\ket{011}$ states (linked by an allowed ESR transition), but not the corresponding states with $\ket{\text{NMR}}=\ket{0}$ (i.e. $\ket{100}$ and $\ket{010}$, linked by a forbidden ESR transition). 

Crucially, we devise our operation to be adiabatic: in a more detailed description included in the next section, we will show that high gate fidelities can be thus achieved across the wide range of $J$ couplings expected from the non-exact positioning of the implanted donors~\cite{oscilla2}, the details of the local electric environment confining the quantum dots, and the noise intrinsic to the control voltages~\cite{electricnoise}. The adiabaticity of this site-selective SWAP, which we will call $\mho$, is maintained if $\hbar \dot{J}\ll \Delta_{0}^{2}$: here, $\Delta_{0}$ corresponds to the initial energy difference between the two states $\ket{101}$ and $\ket{011}$, which is limited by the experimental ability to sweep the dc global magnetic field $B_{0}$. Larger magnetic field excursions would give slower gates: We assume that $\SI{10}{\milli\tesla}$ sweeps can be realized within $\SI{1}{\micro\second}$. It has been demonstrated that the back gate voltage could be switched by the required amount within this time window~\cite{PhysRevB.88.075416}, leading to effective population transfer within a realistic operational framework. 

With the DQ swapped to the ESR states of the selected donors by our operation $\mho$, microwaves can drive a transition on the ESR qubits, conditioned on the state of the NMR qubits. This conditional excitation $\Pi_{\text{ESR}}$ is nothing more than pulsed Electron Nuclear Double Resonance (ENDOR)~\cite{feher}. These donor CNOT fields can be applied globally, since those sites where the SWAP was not done will have their ESR qubit initialized to $\ket{0}$, and the CNOT has no effect -- our surface code implementation does not need any local magnetic field selectivity. The strong hyperfine mixing near the clock transitions in Si:Bi allows the ENDOR transitions to be driven through the electronic part of the states, and thus can be as fast as conventional ESR pulses~\cite{natureactivation}. Following the donor CNOT operation the exchange interaction can be reestablished and the ESR qubit swapped back to the dot electron: the overall result is a donor/dot CNOT with the NMR qubit as control and Dot qubit as target. This gate can be turned into the basic operation required to maintain the surface code, i.e. a CNOT with the data qubit as control and the MQ as target, via the application of four Hadamard gates $\mathcal{H}_{i}$:
\begin{equation}\label{sequence}
\text{Surface code CNOT}=\mathcal{H}_{\text{dot}}\mathcal{H}_{\text{NMR}}\hspace{.3mm}\mho\hspace{.3mm}\Pi_{\text{ESR}}\mho^{-1}\mathcal{H}^{-1}_{\text{dot}}\mathcal{H}^{-1}_{\text{NMR}}.
\end{equation}

After each DQ has been moved to perform a surface code CNOT with its four neighboring donors, the spin state of those donors must be measured and reinitialized, and a similar protocol (without the Hadamard gates in Eq.~7) performed for the X stabilizer measurement. 

The splitting between the allowed and forbidden transitions is only about 2 MHz, while selectively exciting one and not the other is a building block of the CNOT gate. Therefore, if the difference in transition energy provides the only selectivity, the pulses can be no shorter than about 250 ns to avoid exciting the other transition. However, these two transitions are excited by microwaves of opposite helicity, and photon polarization can be used to excite them selectively with short pulses~\cite{PhysRevB.85.094404,PhysRevLett.105.067601}. 

The protected quantum memory designed so far could be readily endowed with defects and braiding, that allow the definition of logical qubits within a surface code~\cite{horsman2012}. To form a defect at any donor site the stabilizer operations are blocked by not applying the voltage to that bottom gate: thus the exchange interaction and SWAP is disabled. The other quantum gates in Eq.~7 either only affect the donor qubits, which are reset anyway by the measurement step, or are pairs of Hadamard gates, which reduce to the identity.

\section{Addressable adiabatic donor/dot SWAP}\label{adiabatic transfer} We now turn to the full characterization of the dynamics that can lead to a robust operation $\mho$ as sketched above: we restrict ourselves to the basis of the four states $\ket{2}, \ket{6}, \ket{4}, \ket{8}$ defined in Eq.~\ref{statesthree}, which correspond to the configuration with NMR qubit fixed to `1', because of the selection rules explained before. 

As shown in Fig.~3(a), we propose initializing the dc magnetic field at a point away from the anticrossing, where the states are not mixed. This is a `quiet' phase configuration at a field of $B_{0}(-t_{0})=B^{\ast}_{0}-\Delta B_{0}$, where $B^{\ast}_{0}$ marks the $E_{4}-E_{6}$ degeneracy point and $\Delta B_{0}\approx \SI{5}{\milli\tesla}$. The field is then swept through the anti crossing and beyond, up to $B_{0}(+t_{0})=B^{\ast}_{0}+\Delta B_{0}$. The coil currents that generate the dc magnetic field could heat the device unacceptably if the sweep rate were too fast, thus we assume that a sweep of $\SI{10}{\milli\tesla}$ is attainable within a $2 t_{0}\approx \SI{2}{\micro\second}$ time interval. This is the fundamental limitation on the speed of the proposed gate.

In the meantime, the exchange coupling is turned on adiabatically from its quiet value $J(-t_{0})$, that is much smaller than the initial detuning $|E_{4}(-t_{0})-E_{6}(-t_{0})|\approx \SI{140}{\mega\hertz}$, to some maximum $J_{\text{max}}$, which is maintained at $t=0$ when $E_{4}(0)=E_{6}(0)$, and then back to the quiet stage [see Fig.~3(a)]. In Appendix B we show that it is sufficient to ramp the voltage up by about $\SI{10}{\milli\volt}$ to increase the exchange coupling by three orders of magnitude for typical device parameters. The qubit states at $t=-t_{0}$ will adiabatically follow the instantaneous eigenstates of the time-dependent evolution, hence if the coupling is strong enough during a sufficient interval of time, a strong population transfer between the diabatic $\ket{4}$ and $\ket{6}$ states will take place. 

More precisely, the time evolution operator induced by the dynamics just outlined in the four-state basis above will lead to the block matrix
\begin{equation}\label{swapwronglarge}
\left(\begin{array}{ccc} e^{-\frac{i}{\hbar} \int\limits_{-t_{0}}^{t_{0}} dt E_{2}}   & 0&0  \\ 0  & e^{-\frac{i}{2\hbar} \int\limits_{-t_{0}}^{t_{0}} dt (E_{6}+E_{4})}U_{-t_{0}}^{t_{0}}  &  0  \\ 0  &  0  & e^{-\frac{i}{\hbar} \int\limits_{-t_{0}}^{t_{0}} dt E_{8}}\\
 \end{array}\right),
\end{equation}
where the eigenergies of the corresponding eigenstates defined in Eq.~\ref{statesthree}
\begin{equation} \begin{array}{l}
E_{2}(t)=E_{-4}^{2-}(t)-B_{0}(t)\frac{\gamma_{e}}{2}-\frac{J(t)}{4}\cos{\theta_{-4}(t)},\\
E_{8}(t)=E_{-3}^{2+}(t)+B_{0}(t)\frac{\gamma_{e}}{2}+\frac{J(t)}{4}\cos{\theta_{-3}(t)},\\
E_{6}(t)=E_{-4}^{2-}(t)+B_{0}(t)\frac{\gamma_{e}}{2}-\frac{J(t)}{4}\cos{\theta_{-4}(t)},\\
E_{4}(t)=E_{-3}^{2-}(t)-B_{0}(t)\frac{\gamma_{e}}{2}-\frac{J(t)}{4}\cos{\theta_{-3}(t)},\\
\end{array}
\end{equation}
combine the energies $E_{m}^{2\pm}$ of the isolated donor states in Eq.~\ref{donoreigen} (fully defined in Appendix A), the Zeeman energy of the dot state and the appropriate exchange coupling. The two-state transfer matrix $U_{-t_{0}}^{t_{0}}$ can be written as 
\begin{equation}\label{swapwrong}
\left(\begin{array}{cc} a(J;t_{0}) \hspace{.5mm}e^{i \psi(J;t_{0})} &  \sqrt{1-a(J;t_{0})^{2}}\hspace{.5mm}e^{i \phi(J;t_{0})}\\  -\sqrt{1-a(J;t_{0})^{2}}\hspace{.5mm}e^{-i \phi(J;t_{0})} & a(J;t_{0}) \hspace{.5mm}e^{-i \psi(J;t_{0})}		\end{array}\right),
\end{equation}
where $a$ is a real number, and $\psi$ and $\phi$ are two real phases: the functional dependence of the propagator on the time profile of the exchange coupling and the duration $t_{0}$ of the pulses has been made explicit. As we calculate in the next section, and show in Fig.~3(b), at any instant $J(t)$ can change by at least two orders of magnitude across all the parallel donor/dot pairs, though it always has the same time dependent profile. An adiabatic evolution where the exchange is pulsed slowly with respect to the magnitude of the initial detuning, $\hbar \dot{J}(t)\ll (\SI{140}{\mega\hertz})^{2}$, allows us to achieve high population transfer fidelities, i.e. to make $a$ small enough, over this large range of $J$ couplings. This way, $U_{-t_{0}}^{t_{0}}$ resembles a SWAP operation over the very wide range of parameters typical of the orbital electron states in a scaled architecture. 

More quantitavely, we have simulated the exact time evolution of the system in the adiabatic regime just defined, with $\Delta(t)=-\Delta_{0}(\frac{t}{t_{0}}), -t_{0} \leq t \leq t_{0}$, $t_{0}=\SI{2}{\micro\second}$ and $J(t)=J_{0} (1-\exp[(|t|-t_{0})/\sigma])$, where $\sigma=\SI{0.9}{\micro\second}$ sets a realistic timescale for tuning the back gate voltage. The fidelities of population transfer $1-a^{2}$ as a function of different donor/dot separations are shown in Fig.~3(b): fidelities higher than $99.9\%$, thus within the $0.1\%$ error rate per operation desired for the surface code with a reasonable overhead~\cite{PhysRevA.86.032324}, correspond to the green region of the plot. Thus almost all of the donor/dot pairs addressed by a local exchange-tuning can undergo a fault-tolerant operation within $\SI{2}{\micro\second}$. This fidelity can thus be maintained with two orders of magnitude variation of the donor-dot exchange coupling. This is a conservative estimate of the calculated range of interactions that the pairs could experience across an array like Fig.~\ref{fig1}.

However, the realization of a Dot/NMR CNOT gate as proposed in Eq.~7 would not follow immediately if the Dot/ESR SWAP $\mho$ were implemented by the time evolution defined in Eqs.~\ref{swapwronglarge} and ~\ref{swapwrong}. The reason for such failure lies in the presence of the $J$-dependent phase $\phi\neq 0$, which implies that the operator in Eq.~\ref{swapwrong} has entangling power: its action would not be limited to SWAPping the quantum states involved. This problem is solved by the sequence illustrated in Fig.~3(a), that combines the time evolution in Eq.~\ref{swapwrong} with a `phase-erasing' operation. The only extra ingredients required by this recipe are selective Rabi resonant pulses that could be achieved with high fidelity for $\SI{}{\micro\second}$ gating times, followed by extra adiabatic tuning sequences of magnetic field and back gate voltage. The propagator of this updated sequence is, up to an irrelevant multiplicative phase, 
\begin{equation}\label{gatemio}
\mho=\left(\begin{array}{cccc} e^{-\frac{i}{\hbar}\xi } & 0 & 0 & 0 \\ 0 & a\hspace{1mm} e^{i \theta} & -\sqrt{1-a^{2}} & 0 \\ 0 & \sqrt{1-a^{2}} & a\hspace{1mm} e^{-i \theta} & 0 \\ 0 & 0 & 0 & e^{-\frac{i}{\hbar}\xi}\\
 \end{array}\right),
\end{equation}
where the phase $\phi$ has now disappeared, and $\xi=\int\limits_{-t_{0}}^{t_{0}} dt (E_{2}+E_{8}-E_{6}-E_{4})$. Straightforward matrix multiplication shows that the complete sequence in Eq.~7, assuming for simplicity a perfect Dot/ESR transfer ($a=0$), leads to the following time propagator in the complete basis of Eq.~\ref{statesthree}:
\begin{equation}
-i\left(\begin{array}{cccc} \mathcal{I} & 0 & 0 & 0 \\ 0 & \begin{array}{|cc|}\cline{1-2}0 & 1\\ 1 & 0\\ \cline{1-2}\end{array} & 0 & 0 \\ 0 & 0 & \begin{array}{|cc|}\cline{1-2}0 & 1 \\ 1 & 0 \\ \cline{1-2}\end{array} & 0 \\ 0 & 0 & 0 & \mathcal{I}\\
 \end{array}\right).
\end{equation} 
This is seen to coincide with the desired surface code CNOT, when restricted to the degrees of freedom that effectively host the data and measurement qubits, namely the states with the ESR qubit being initialized to `0': $\ket{1}, \ket{2}, \ket{5}, \ket{6}$. Crucially, this form of the time propagator is retained within the surface code error tolerance across almost all donor/dot pairs in a realistic scaled donor/dot computer. 
\begin{figure}[t!]
  \centering
\begin{minipage}{.49\textwidth}
\begin{overpic}[width=.99\textwidth]{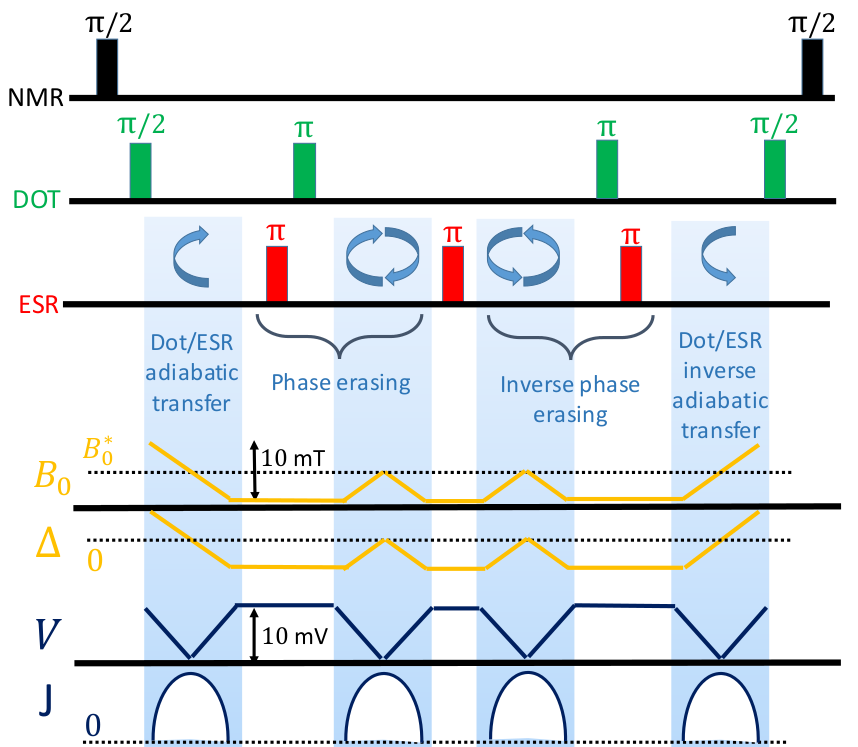}
 \put (55,80) {\Large{$(a)$}}
\end{overpic}
\end{minipage}
\vspace{.3cm}
\\
\begin{minipage}{.49\textwidth}
\begin{overpic}[width=.99\textwidth]{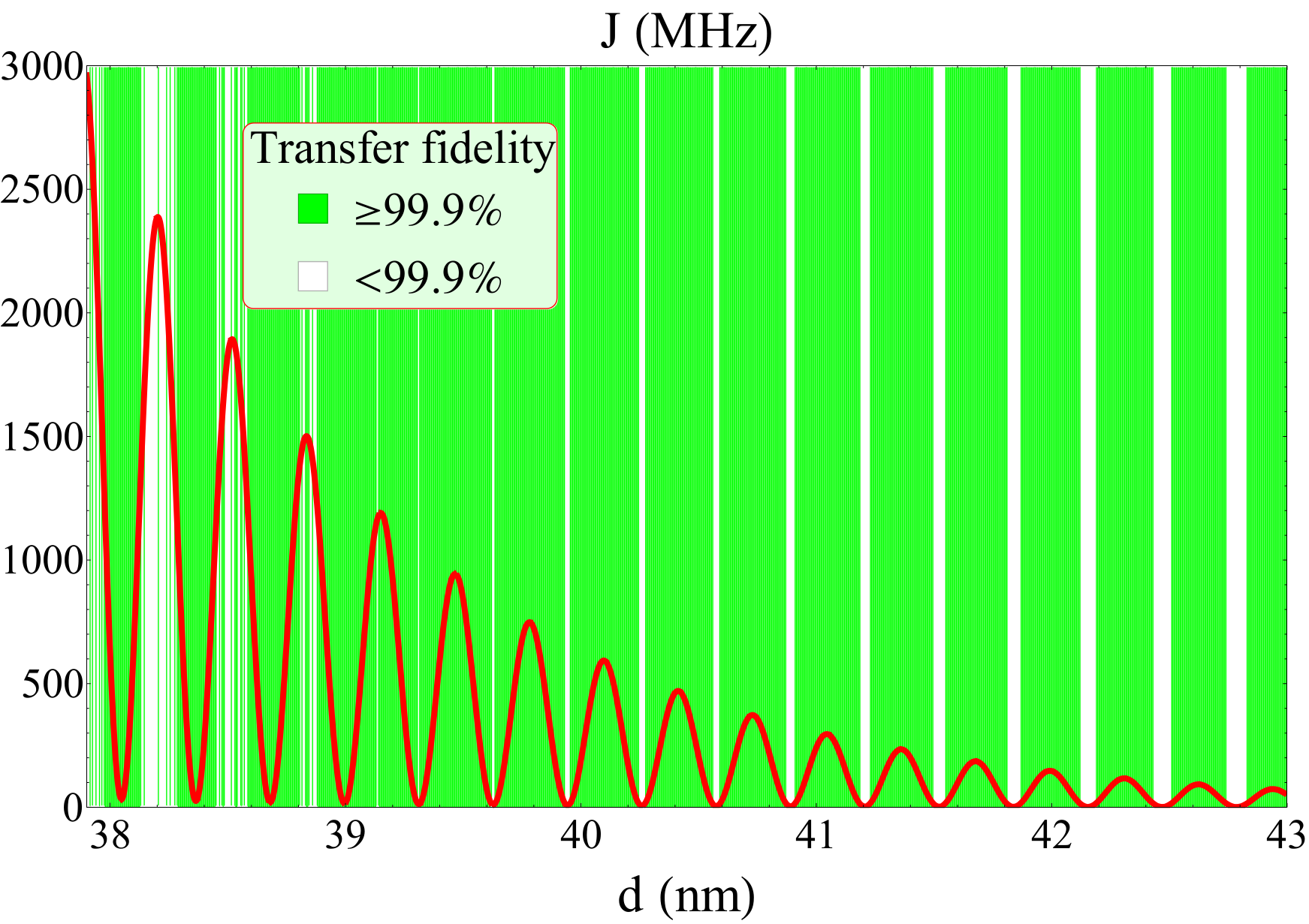}
 \put (55,58) {\Large{$(b)$}}
\end{overpic}
\end{minipage}
\caption{$(a)$ Complete CNOT sequence between the Dot and the NMR qubit, that includes performing resonant Hadamard ($\pi/2$) and $\pi$ Rabi pulses on the different qubits, plus adiabatic Dot/ESR transfer sequences $\mho$ (light blue boxes) based on linear dc ramping of the magnetic field $B_{0}$ and the back-gate voltage, $V$. The adiabatic transfers take place thanks to the $J$ coupling turned on at the crossing points of the Dot/ESR levels ($\Delta=0$), while `phase erasing' steps allow the final quantum state to be independent of the evolution induced by each particular $J$. All operations require a few $\SI{}{\micro\second}$. $(b)$ The red line shows a best fit to a set of calculations of exchange splittings that the different donor-dot pairs of electrons would experience within a scaled architecture, due to the imprecise implantation depth $d$ of the donors (an equivalent applied electric field $F=\SI{4}{\kilo\volt/\centi\meter}$ is assumed). The pairs within the green regions will undergo adiabatic population transfer with fidelity higher than $99.9\%$. These regions become thinner to the left edge of the plot, where the high couplings break the adiabaticity condition, and to the right edge, where the couplings are too weak to induce an efficient depopulation at the avoided crossing. }
\label{fig3}
\end{figure}

\section{Strong variability of qubit couplings and splittings within a scaled architecture} Order-of-magnitude oscillations in $J$ are expected as the position of the donor changes with respect to the abrupt heterointerface where the quantum dot is formed. This feature is intrinsic to the silicon band structure~\cite{bhatt,oscilla2,PhysRevB.89.235306}, and makes high-fidelity hard to attain with the dynamical phase gates proposed in donor-based quantum computers proposals~\cite{kane98}. In Fig.~3(b) we show the results of our multi-valley effective mass theory calculation of the exchange energy of a donor and dot electron as a function of the distance from the surface to the donor ion, assuming a field of 4 kV/cm is confining the dot electron. This theoretical approach has been introduced in Refs.~\onlinecite{PhysRevB.89.235306,PhysRevB.90.195204}, and the details of how it has been adapted to the system considered here are given in Appendix B. It is clear that moving the donor one lattice constant closer or farther from the surface can bring the exchange interaction from a peak to a trough. The placement of phosphorus donors in a single plane has been demonstrated using hydrogen lithography on Si~\cite{Fuechsle2012}, but an analogous approach for Bi donors has not been developed. Placement of Bi donors by ion implantation is associated with larger alignment uncertainties~\cite{Schenkel2013}. 

Furthermore, the exchange coupling between any two confined electron spins in silicon is known to depend exponentially on the magnitude of a uniform applied electric field, as we have confirmed in Appendix B for a donor/dot pair. The only alternative to high precision, individual tuning of gate voltages at each qubit site is to develop, as we have proposed, manipulations that possess an innate insensitivity to these orbital details.

It is expected that non-Markovian noise on the applied control voltages, which represents another major hurdle to Loss-DiVincenzo two-qubit gates~\cite{electricnoise}, will also be effectively combated by such insensitivity -- although a more complete analysis is a subject for future work. In Appendix C we show how our surface code CNOT is also extremely robust against local magnetic noise that affects both the donor and the dot spins.

\section{Conclusions} In summary, we have shown how bismuth donors and quantum dots in a silicon host can be combined into a surface code quantum computer architecture. The first insight of this scheme is that coherent donor spins can be positioned microns apart, which is compatible with the current state of the art in silicon fabrication, as connections between neighboring donors are mediated by the quantum dot electrons. This requires a robust, addressable way of SWAPping information from a donor to a dot and vice-versa, which we construct from a NMR-controlled Fredkin gate between the donor and the dot electrons coupled by the exchange interaction. In contrast to previous proposals of two-qubit gates built on dynamical exchange phases, we devise adiabatic manipulations that are insensitive to a two order-of-magnitude variation in the interaction strength. Combining this with high-fidelity microwave driven ENDOR transitions on the highly coherent bismuth donors, we construct CNOT gates for surface code error correction that retain high fidelity without the need for individual tuning of the orbital electron states. Moreover, all microwave fields are applied globally, and qubits are locally selected for an operation by switching the back-gate voltage to control the donor/dot SWAP. As the nearest-neighbor coupling required to implement the surface code could be achieved by shuttling the array of electrons in unison with CCD-like gates, the feasibility of the local control needed is greatly improved as compared to previous blueprints. With micron pitch structures the surface code would have ${10}^{8}$ physical qubits per square centimeter, allowing for many error-corrected logical qubits. The insensitivity to donor-dot alignment variations (see Appendix B) may enable fabrication of large donor-dot arrays by ion implantation~\cite{Schenkel2013}.

\begin{acknowledgements} This research was funded by the joint EPSRC (EP/I035536)/ NSF~(DMR-1107606) Materials World Network grant (GP, BWL, SAL), the EPSRC grant EP/K025562 (BWL), the NSF MRSEC grant DMR-01420541 (SAL), and the Department of Energy, Office of Basic Energy Sciences grant DE-SC0002140 (RNB) and the DOE Office of Science under contract no.~DE-AC02-05CH11231 (TS). GP thanks the University of St Andrews and EPSRC for a Doctoral Prize Fellowship.
\end{acknowledgements}

\section*{Appendix A: Choice of donor}
\label{donor}
Si:Bi systems have now been experimentally established as excellent candidate qubits~\cite{PhysRevLett.105.067601}. Bi 9/2 nuclear spins combined with the 1/2 donor electron spins provide a rich Hilbert space of states from which to choose the qubit logical $|0\rangle$ and $|1\rangle$. Their hyperfine interaction is the strongest available among the group V substitutional donors in Si, which makes it easier to transfer the information from the electron to the nuclear spin; moreover, it allows the existence of so-called clock transitions~\cite{clocktransitions}, i.e. transitions between hyperfine mixed nuclear-electron spin states that are very insensitive to the actual magnetic field of the environment. We will now identify the specific donor states that host the measurement qubits that we propose to couple to the dot data qubits within the surface code.\\
The mixed Hilbert space set up by the electron and nuclear spin levels of a group V donor is governed, in the presence of a fixed magnetic field $\textbf{B}_{0}$, by the Hamiltonian
\begin{equation}\label{levels}
H=\gamma_{e} \textbf{B}_{0}\cdot\textbf{S}-\gamma_{n} \textbf{B}_{0}\cdot\textbf{I}+A\hspace{.5mm} \textbf{S}\cdot\textbf{I},
\end{equation} 
where $A$ is the hyperfine interaction between the nuclear spin $\textbf{I}$ and the electron spin $\textbf{S}$, $\gamma_{e}=\dfrac{g_{e} \mu_{B}}{\hbar}=\SI{27.997}{\giga\hertz/\tesla}$ is the magnetic moment of the electron, $\gamma_{n}=\SI{0.007}{\giga\hertz/\tesla}$ is the nuclear magnetic moment.\\
The behaviour of the corresponding spectrum in the region of intermediate $\textbf{B}_{0}$ (roughly speaking, when $ A\approx \gamma_{e} B_{0}$) can get very interesting, if $I$ is large enough. In fact, apart from Si:P, the manifold of the mixed levels for all group V donors (As, ${}^{111}$Sb, ${}^{113}$Sb, and Bi with respective $I$=3/2, 5/2, 7/2, 9/2) allows for specific values of $B_{0}$, where the energy difference $f$ between selected mixed eigenstates has a minimum, i.e. $\dfrac{\partial{f}}{\partial B_{0}}=0$~\cite{PhysRevB.85.094404}. The immediate and useful consequence of such rich behaviour, conceptually due to the large number of mixed levels available, is that the $T_{2}$ of a qubit stored in the two donor levels separated by a `clock transition' will not suffer from local fluctuations in the magnetic field, which include hyperfine and dipolar interaction with the ${}^{29}$Si nuclei and paramagnetic coupling to other electrons and impurities.\\
The eigenenergies corresponding to the states of the isolated donor in Eq.~\ref{levels} are
\begin{align}\nonumber \label{energies2m}
E^{2 \pm}_{m}= & -\frac{A}{4}-B_{0}\gamma_{n} m \hspace{1mm} \pm  \\ 
& \sqrt{A^{2} [I(I+1)+\frac{1}{4}-m^{2}]+(A m+B_{0} \gamma_{e}+B_{0} \gamma_{n})^{2}}, &   
\end{align}
for $-I-1/2<m<I+1/2$, and
\begin{equation} 
E^{1 \pm}_{m}= \pm \frac{1}{2}(A m+B_{0} \gamma_{e}+B_{0} \gamma_{n})-\frac{1}{4}(A+4 B_{0} \gamma_{n} m ),
\end{equation}
for $m=\pm (I+1/2)$. \\
In a regime of intermediate $\textbf{B}_{0}$ values, large mixing between $|m_{S}\rangle$ and $|m_{I}\rangle$ states ensues that allowed $\SI{}{\giga\hertz}$ transitions occur between states of the form $|\pm, m\rangle \leftrightarrow |\pm, m-1\rangle$ and $|\pm, m\rangle \leftrightarrow |\mp, m-1\rangle$. It has been observed experimentally~\cite{PhysRevLett.105.067602} and then clarified theoretically~\cite{PhysRevB.85.094404} that the sweet spots aforementioned occur for the second kind of transitions, namely when:
\begin{equation}\label{clockspot}
B_{0}=B_{0}^{\ast}\approx -\frac{A}{\gamma_{e}} \frac{(m-1) g(m)+ m g(m-1)}{g(m)+g(m-1)},
\end{equation}
with the restriction -$I$+3/2 $ \leq m \leq 0$, where $g(m)\equiv \sqrt{I(I+1)+\frac{1}{4}-m^{2}} $. The nature of the expression \ref{clockspot} should clarify why Si:P does not show any clock transition, while their number increases for larger $I$, as there will be more integers $m\leq 0$ able to satisfy such condition. \\
Since we suggest the `hybridization' of a highly coherent donor system with the two Zeeman split states of a quantum dot electron spin, we would like the Zeeman dot frequency $f_{\text{dot}}\approx B_{0} \gamma_{e}$ (the dot electron g-factor is 1.997) to cross some donor clock transition $f_{\text{donor}}$ at the sweet spot where $\dfrac{\partial f_{\text{donor}}}{\partial B_{0}}=0$, as shown in Fig.~2. Let us show that meeting those requirements automatically selects Si:Bi as the only option among group V donors in silicon: after rearranging Eq.~\ref{clockspot}, we get
\begin{equation}\label{clockreq}
f_{\text{dot}}=\gamma_{e} B_{0}^{\ast}\approx - A \left[ m - \left(1+\dfrac{2 m-1}{g(m)^2}\right)^{-1/2}\right]<-A (m-1).
\end{equation}  
On the other hand, close to a clock transition, the eigenenergies of the hybrid states involved can be approximated as
\begin{equation}
E^{2 \pm}_{m}\approx  -\frac{A}{4} \pm \frac{A}{2} \sqrt{I(I+1)+\frac{1}{4}-m^{2}}.
\end{equation}
Hence, requiring that a $|\pm, m\rangle \leftrightarrow |\mp, m-1\rangle$ transition is degenerate with $f_{\text{dot}}$ implies 
\begin{equation}
f_{\text{donor}}-f_{\text{dot}}= \frac{A}{2}\left[ g(m)+g(m-1)\right]-\gamma_{e} B_{0}^{\ast}=0 .
\end{equation}
By virtue of \ref{clockreq},
\begin{equation}
f_{\text{donor}}-f_{\text{dot}}>\frac{A}{2}\left[g(m)+g(m-1)+2(m-1)\right], 
\end{equation}
and this latter expression is seen to be always positive unless $m\leq-3$. Thus the required degeneracy can be achieved only within Si:Bi, specifically addressing the $|+, m=-3\rangle \rightarrow |-, m=-4\rangle$ transition, which we call allowed transition, and the $|+, m=-4\rangle \rightarrow |-, m=-3\rangle$ transition, the forbidden one. The following hybrid electron-nuclear spin states are respectively involved (the colored arrows label the qubit transitions as indicated in Fig.~3):  
\begin{equation}\begin{array}{c}\label{donorstates}
\cos{\dfrac{\theta_{-3}}{2}}|1/2, -7/2 \rangle + \sin{\dfrac{\theta_{-3}}{2}}\ket{-1/2, -5/2} \\
\normalsize f_{\text{donor}}=\SI{5.2142}{\giga\hertz}  \hspace{3mm}\large \textcolor{red}{\mathbf{\downarrow}}   \hspace{3mm} \normalsize B_{0}^{\ast}=\SI{0.188179}{\tesla} \\ -\sin{\dfrac{\theta_{-4}}{2}}|1/2, -9/2 \rangle + \cos{\dfrac{\theta_{-4}}{2}}|-1/2, -7/2 \rangle,\\
\\
-\sin{\dfrac{\theta_{-3}}{2}}|1/2, -7/2 \rangle + \cos{\dfrac{\theta_{-3}}{2}}\ket{-1/2, -5/2} \\
\normalsize f_{\text{donor}}=\SI{5.21683 }{\giga\hertz} \hspace{3mm}\large \textcolor{blue}{\mathbf{\uparrow}} \hspace{3mm }\normalsize B_{0}^{\ast}=\SI{0.188086}{\tesla} \\
\cos{\dfrac{\theta_{-4}}{2}}|1/2, -9/2 \rangle + \sin{\dfrac{\theta_{-4}}{2}}|-1/2, -7/2 \rangle.
\end{array}
\end{equation}
The occurrence of an energy crossing between the Zeeman transition linking the two dot electron spin states and the allowed donor clock transition is displayed in Fig.~2(a). In the high field limit (Zeeman much larger than hyperfine) the right transition in Eq.~\ref{donorstates} is forbidden, since it involves a nuclear spin flip: this is the motivation for labeling the left transitions as `allowed' and the right as `forbidden', even though both are actually enabled in the intermediate $B_{0}$ regime investigated here. Each of these two transitions couples to opposite helicity microwave photons, as noted by Ref.~\onlinecite{natureactivation}, thus the selective excitation of a single transition in the pair does not pose fundamental physical limitations, even though the energy difference between the two, about $\SI{2}{\mega\hertz}$, would hardly be distinguished with fast microwave pulses.

\section*{Appendix B: Exchange coupling between a MOS quantum dot and a Si:Bi donor} 
\begin{figure}[h!]
\centering
   \includegraphics[width=.5\textwidth]{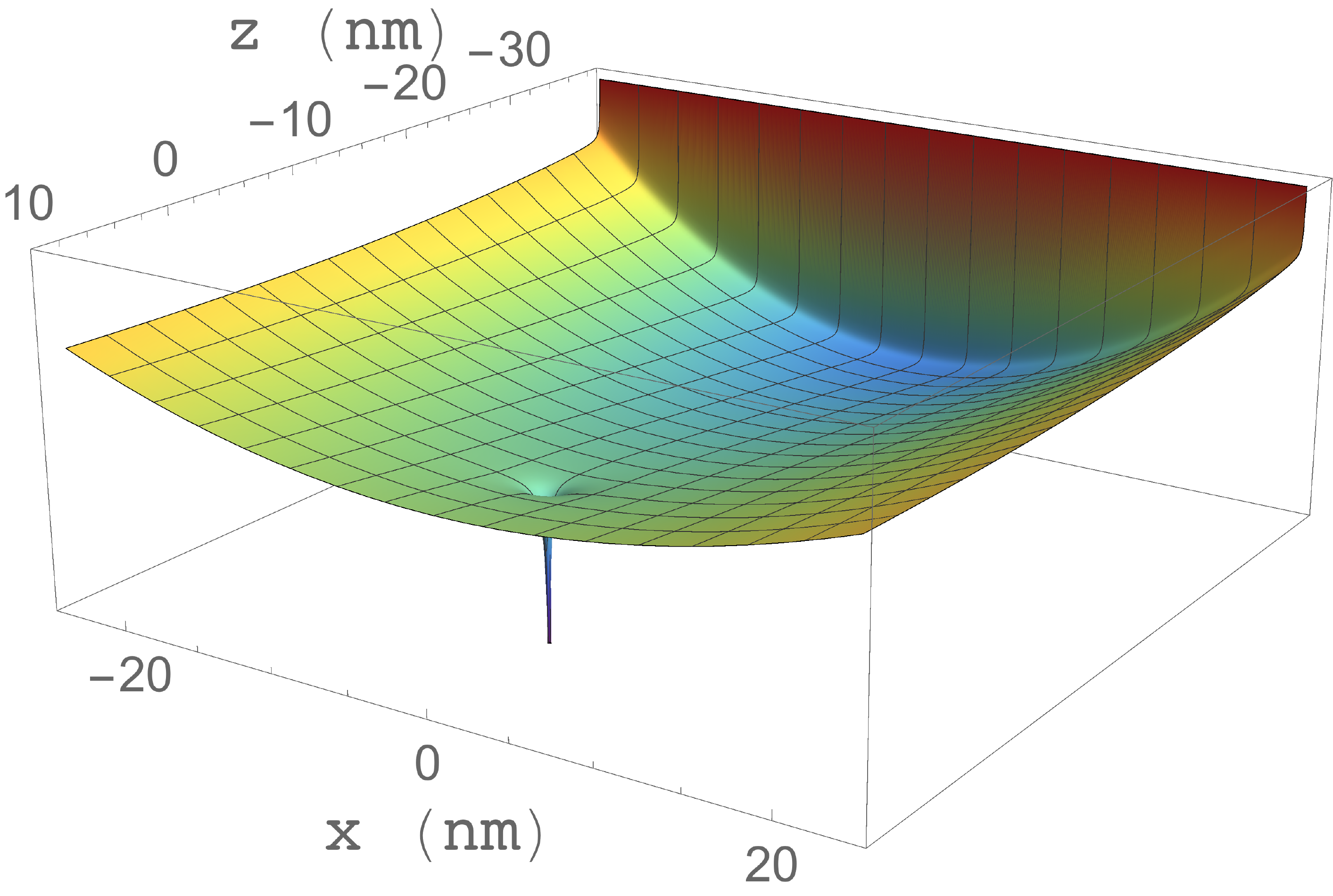}
  \caption{A three dimensional plot of the two-well potential in Eq.~\ref{potential}, that binds the quantum dot electron at the interface with the oxide and the donor electron in the region close to the implanted dopant Bi atom, in the $x-z$ ($y=0$) plane. The origin of our coordinate system resides at the position of the Bi nucleus, while $z=-d$ corresponds to the interface plane. An electric field $F=\SI{20}{\kilo\volt/\centi\meter}$ and a donor depth $d=\SI{38}{\nano\meter}$ are assumed.}
  \label{fig:potential}
\end{figure}

\begin{figure}[t!]
\centering
    \includegraphics[width=.5\textwidth]{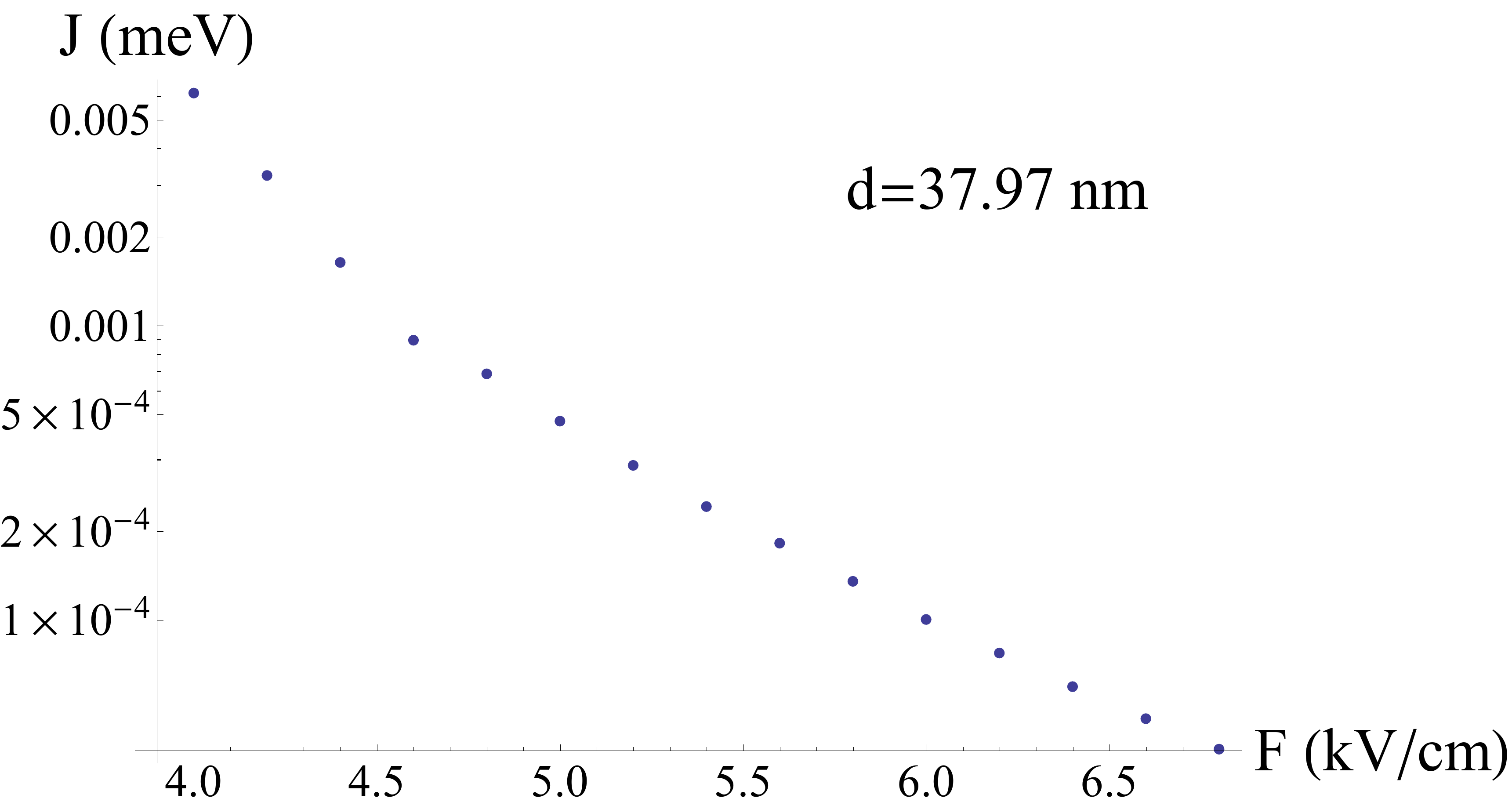}
  \caption{Donor/dot exchange splittings as a function of the applied field: as $F$ increases the dot is more localized at the interface, thus the interaction decreases. The control is very efficient: tuning the field by less than 3 kV/cm allows one to switch $J$ `off' by two orders of magnitude. A donor depth of $d=37.97$ nm is assumed, but the same trend would be followed for any position of the Bi nucleus. At smaller fields the influence from the Coulomb attraction from the Bi impurity is still significant, and it affects the confinement of the dot state; then, from $F\gtrsim 4.5$ kV/cm, $J$ becomes relatively less sensitive to the applied field, as the interface well is now more strongly established.}
  \label{fig:F}
\end{figure}

The aim of this section is to evaluate the exchange coupling that would arise between an electron spin which is confined in a quantum dot close to a Si/SiO$_{2}$ interface and the excess electron spin provided by a donor Bi atom implanted deep in the bulk of a Si layer, at a distance $d$ from the interface. This interaction paves the way for the fundamental data-measurement qubit coupling that is needed for the surface code proposed in the Results.\\
The confinement for the interface electron is provided by an external electric field $F$ (in the $\hat{z}$ direction, which we assume to be perpendicular to the plane which contains the interface) and by a quantum dot potential (approximately parabolic) in the transverse $x-y$ plane. This simple modeling accounts for the voltage landscape that the confining interface gates would be able to produce. The impurity potential due to the substitutional implanted Bi atom completes the description of this two-electron problem: the potential energy of an electron in this system, as shown in Fig.~\ref{fig:potential}, is described as:
\begin{equation}\label{potential}
V(\textbf{r})=+ e F z + U_{\text{dot}}(\textbf{r}) + U_{\text{donor}}(\textbf{r}) + U_{\text{image}}(\textbf{r}),
\end{equation}
where $\rho$ is the radial coordinate in the plane of the interface, $U_{\text{dot}}(\textbf{r})=\frac{\omega_{0}}{2}\rho^{2}$ is the confining potential of the dot gates, $U_{\text{donor}}(\textbf{r})=- \frac{e^{2}}{\epsilon_{Si}\sqrt{\rho^{2}+z^{2}}}(1-e^{-b \sqrt{\rho^{2}+z^{2}}}+B \sqrt{\rho^{2}+z^{2}}e^{-b \sqrt{\rho^{2}+z^{2}}})$ represents the the Si:Bi impurity potential outside the donor central cell~\cite{PhysRevB.90.195204} (with $\epsilon_{Si}=11.4$ is the dielectric constant of Silicon, $b$ and $B$ two pseudopotential parameters), $U_{\text{image}}(\textbf{r})=\frac{e^{2} Q}{\epsilon_{Si}\sqrt{\rho^{2}+(z+2 d)^{2}}}-\frac{e^{2}Q}{4\epsilon_{Si}(z+d)}$ parametrizes the electrostatic image effects due to the dielectric barrier, with
$Q=\dfrac{\epsilon_{SiO_{2}}-\epsilon_{Si}}{\epsilon_{SiO_{2}}+\epsilon_{Si}}$ and $d$ the distance of the nucleus from the interface. 
The infinite wall at the interface models the $\approx \SI{3}{\electronvolt}$ step between the energies of the conduction band edges of the silicon and the oxide layer, and implements our assumption that the dot electron state does not penetrate significantly into the oxide. The electric field is assumed to be uniform and unidirectional throughout the system, which is a reasonable approximation for realistic devices of this kind, as is the parabolic transverse confinement that binds the quantum dot. We take into account the effect of the accumulation of charges on the dielectric SiO$_{2}$ boundary, induced by the electrostatic configuration in the Si layer, via the image-charge method~\cite{externalexchange}.\\
Effective mass theory is used in our evaluations of exchange splittings, since the latter will gain the most relevant contributions from the electronic densities in the intermediate spatial region between the two wells, i.e. far from the Bi nuclear cell where EMT fails. Our theory for the donor state has been tested in Ref.~\onlinecite{PhysRevB.90.195204} to yield good agreement with experimental Stark shifts of the spin Si:Bi spectrum. The donor wavefunction is there given by
\begin{equation}\label{donor}
\psi_{D}=\sqrt{\frac{2}{3}}\sum_{i=x,y,z} F^{D}_{i}(\textbf{r}) \cos(\textbf{k}_{0 i}\cdot \textbf{r}) u_{i}(\textbf{r}),
\end{equation} 
where the functions $u_{i}(\textbf{r})$ are the lattice periodic parts of the Bloch eigenstates of the undoped silicon layer relative to each conduction band minimum $\textbf{k}_{0 i}$, and the anisotropic envelopes $F^{D}$ are defined e.g. as:
\begin{eqnarray}\label{newtrial}
\begin{aligned}
F^{D}_{z}= N_{D} \left[\text{e}^{-\sqrt{\frac{x^{2}+y^{2}}{a^{2}_{s}}+\frac{z^{2}}{b^{2}_{s}}}} + \beta \hspace{1mm} \text{e}^{-\sqrt{\frac{x^{2}+y^{2}}{a^{2}_{l}}+\frac{z^{2}}{b^{2}_{l}}}}\right],% \\
%F^{D}_{y}= N_{D} \left[\text{e}^{-\sqrt{\frac{x^{2}+z^{2}}{a^{2}_{s}}+\frac{y^{2}}{b^{2}_{s}}}} + \beta \hspace{1mm} \text{e}^{-\sqrt{\frac{x^{2}+z^{2}}{a^{2}_{l}}+\frac{y^{2}}{b^{2}_{l}}}}\right], \\
%F^{D}_{x}= N_{D} \left[\text{e}^{-\sqrt{\frac{z^{2}+y^{2}}{a^{2}_{s}}+\frac{x^{2}}{b^{2}_{s}}}} + \beta \hspace{1mm} \text{e}^{-\sqrt{\frac{z^{2}+y^{2}}{a^{2}_{l}}+\frac{x^{2}}{b^{2}_{l}}}}\right],
\end{aligned}
\end{eqnarray}    
with different pairs of Bohr radii distinguishing the short ($a_{s},b_{s}$) from the long ($a_{l},b_{l}$) range hydrogen-like decay, with a relative weight $\beta$ ($N_{D}$ is a normalization factor). In the regime of donor depths and electric fields of interest here,
the donor state can be assumed, to very high precision, to coincide completely with the bulk ground eigenstate, which is constructed from an equal superposition of the Bloch functions of all the six degenerate valleys. \\
The issues due to the valley degeneracy of the silicon conduction band are completely taken into account for the donor state, while we assume that the interface state resides in only one of the two $\hat{z}$-valleys combinations (namely, the symmetric one) that are almost degenerate close to the interface. Such degeneracy is known to be removed by the $z$-confinement provided by the Si/SiO$_{2}$ boundary and the electric field~\cite{PhysRevLett.96.096802}, with splittings as large as $\approx 1$ meV that increase linearly with the applied field $F$~\cite{PhysRevB.84.155320}, but a complete theory of the interplay of those effects will depend crucially on the details of the device. However, since the inter-valley coupling at the interface is not as strong as for a bulk donor, a more refined description would only provide the correct superposition of the two valleys that constitutes the orbital interface ground state, something that will not change qualitatively the analysis below. In fact, our calculations provide a worst case scenario, that is well suited to the feasibility study we are aiming at: the oscillations in $J(d)$ are maximal if the orbital state is an equal weight superposition of $z$ valleys, so that the spatial dependence of the dot wavefunction is exactly in (anti-)phase with the donor one. Due to the roughness of the interface, for example, it is likely that other combinations of the two valleys, with different weights, correspond to the actual dot ground state: out-of-phase valley interference would then be able to reduce the large oscillations calculated here.  \\
The envelope of the dot electron wavefunction is calculated via a variational optimization of its on-site ground binding energy, as determined by the potential in Eq.~\ref{potential}. Based on the strong similarity of the interface well with the exact solvable problem of an infinite triangular well, it has been proposed that a good \emph{ansatz} for the interface envelope should resemble an Airy function~\cite{externalexchange} along the $z$ axis, while a Gaussian confinement is well suited to the $x-y$ confinement:
\begin{eqnarray}\label{interface}\nonumber
&\psi_{I}=\sqrt{2} \cos[{k_{0}(z+d)}] \hspace{1mm}u_{\textbf{k}_{0z}}(\textbf{r}) F^{I}_{z}(\textbf{r}), \\&F^{I}_{z}(\textbf{r})= N_{I} (z+d)^{2}e^{-{\alpha(z+d)}/{2}}e^{-\beta^{2}\rho^{2}/2},
\end{eqnarray}
where $5/\alpha$ gives the typical spread of the wavefunction in the $\hat{z}$ direction, while $2/\beta$ represents its extent in $x-y$ plane ($N_{I}$ is a normalization factor). We solve variationally for the ground eigenstate and eigenvalue by optimizing $\alpha$ and $\beta$ as a function of $F$ and the donor depth $d$. In fact, even if the donor is implanted as deep as $\sim \SI{40}{\nano\meter}$ from the interface, the screened Coulomb attraction from the Bi nucleus affects the dot state in a non-negligible way. It provides a strong enough binding in the (001) plane that the transverse extent of the dot electron amounts to a radius of $\approx \SI{25}{\nano\meter}$, which already matches the length scales of experimental quantum dot engineering. As a consequence, $\omega_{0}$ is neglected in our calculations. \\
The range of magnitudes of the exchange splittings that we need sets rather stringent requirements: if the dot is not tightly confined ($E_{I}\approx \SI{-12}{\milli\electronvolt}$, F$\approx \SI{4}{\kilo\volt/\centi\meter}$), then the donor should be as deep as $\sim \SI{40}{\nano\meter}$. If larger voltage gates are established, then the donors should be positioned closer to the interface, which is harder to realize, and the more efficient hybridization between donor and dot states combined with higher fields will make donor ionization more likely. \\
We use the Heitler-London method~\cite{PhysRevB.89.235306} to calculate the difference between the two lowest eigenvalues of the double electron problem. Within this scheme, the states are the orthonormalized symmetric and antisymmetric orbital superpositions of the product of two single electron functions $\psi_{I}$ and $\psi_{D}$:
\begin{equation}
\Psi(\textbf{r}_{1},\textbf{r}_{2})_{\pm}=\frac{1}{\sqrt{2(1\pm S^{2})}} (\psi^{D}(\textbf{r}_{1})\psi^{I}(\textbf{r}_{2}) \pm \psi^{D}(\textbf{r}_{2})\psi^{I}(\textbf{r}_{1})).
\end{equation}    
Here, $\textbf{r}_{1},\textbf{r}_{2}$ are the spatial coordinates of electron 1 and 2, respectively, and $S=\langle \psi^{D}|\psi^{I}\rangle$ is the overlap of the single-electron ground states. Exchange splittings are plotted in Fig.~4 in the main text as a function of the donor depth, and in Fig.~\ref{fig:F} as a function of the applied field $F$.

We use high precision numerical calculations to estimate the highly oscillatory integrals involved in the calculation of the exchange coupling. The largest ratio between each $J(d)$ maximum and the next closest minimum is $\approx 0.01$ for all the implantation depths considered here. The state-of-the-art implantation processes for donors in silicon allow a precision of $\approx 1$ nm in the depth of the impurities (for example, a very shallow implant followed by low-temperature overgrowth~\cite{Kane2000}), which would correspond to several oscillations, and an estimated relative spread of maximum to minimum $J$ values of 1:200. We remark that a different crystallographic direction could be chosen for engineering the quantum dots, which could reduce the strength of the oscillations in $J(d)$: for example, if the donor and the dot were separated along one [011] axis, then the interface ground state would be a combination of $y$ and $z$ valleys, but only the $F_{z}$ components of the exchange would oscillate with $d$. However, the two-fold valley degeneracy discussed before would then include more states. While the degeneracies would be very likely broken by the confinement and the interface roughness, the dot state would nonetheless be more liable to couple to excited orbital states, which would cause information leakage. \\
Let us highlight that the $J$ values presented here would be completely robust against small displacements of the nominal donor position in the plane transverse to the donor/dot separation: no extra oscillation would take place if the donor and the dot are not completely aligned vertically, since the interface state is only made up of $z$ valleys. This feature contrasts the behaviour of the exchange coupling between two neighbouring donors examined in Ref.~\onlinecite{PhysRevB.89.235306}, where all the valleys contribute to the interference, and thus $J$ is sensitive to displacements along any spatial direction.   \\

\section*{Appendix C: Robustness of the CNOT to local shifts of the qubit frequencies} 
So far we have assumed that all the physical qubits experience the same local magnetic field. However, in a realistic device the resonant frequencies of donor spins will be modified by local shifts of their hyperfine coupling, due to uncontrolled local strain and inhomogeneities of the electrostatic environment~\cite{PhysRevB.90.195204}, while the resonant frequencies of electron spins in quantum dots will be affected by the local spin-orbit interaction.  \\
Typical linewidths of Sb donor spins implanted near a surface are less than $\SI{600}{\kilo\hertz}$~\cite{prb82}. The absolute shifts in the hyperfine coupling of bulk Si:Bi spins due to inhomogeneous electrostatic environments are comparable to those of Si:Sb donors: while the relative sensitivity of the hyperfine coupling for Si:Bi donors is one order of magnitude smaller than for Si:Sb, the unshifted Si:Bi hyperfine coupling $A_{0}$ itself is one order of magnitude larger than Si:Sb~\cite{PhysRevB.90.195204}. Thus $\SI{600}{\kilo\hertz}$ is a good measure of the typical local frequency shift that the Si:Bi donors could experience in the scaled architecture proposed here. When the static magnetic field $B_{0}$ is perpendicular to the surface, as described here, natural quantum dots close to a MOS interface also have a linewidth of less than $\SI{600}{\kilo\hertz}$~\cite{applphys}.
The swap gate described here will not be impeded significantly by such differences across a scaled device: In the first part of the adiabatic time evolution the detuning $\Delta$ is of the order of $\SI{100}{\mega\hertz}$ for all the donor/dot pairs, thus local differences of the order of $\SI{600}{\kilo\hertz}$ would only produce an overall relative shift of the corresponding detuning that is no larger than $1\%$. When, closer to the degeneracy point, the exchange coupling $J$ becomes the relevant energy scale, for the pairs that yield $>$$99.9\%$ transfer fidelity (as described in Fig.~3(a)) $J$ lies in the $\SI{30}{\mega\hertz}$-$\SI{3}{\giga\hertz}$ range, which is again at least $50$ times larger than the frequency shift. The Landau-Zener physics that governs the time evolution leading to the swap operations described above is affected only slightly by such local distortions. More quantitavely, we have verified that simulated transfer fidelities higher than $99.9\%$ are still expected within the same range of donor/dot distances considered in Fig.~3(a), if the local detuning pertaining to each donor depth is shifted by as much as $\SI{10}{\mega\hertz}$ from the reference detuning. Only a negligible fraction of the fault-tolerant donor/dot pairs within the non-shifted array discussed before undergoes adiabatic transfers with fidelities below the $99.9\%$ threshold.

\section*{Appendix D: Layout and operational considerations}

For clarity, in the main text we have considered one particular mode of operation and arrangement of donors and dots, but there are a number of other similar architectures which have advantages and disadvantages.  Here we discuss some of these other possibilities, as well as our current understanding of their strengths and weaknesses.

One possibility would be to avoid the exchange interaction entirely.   Ionizing the donors and moving their electrons to neighboring donors has been suggested as an alternative to exchange-based entanglement of nuclear spins~\cite{Kane2000,morton2009}. However, with hyperfine interaction frequencies of order 100 MHz or higher, controlling the timing of the electron removal and reintroduction to the donors becomes problematic.

In  Fig.~1 of the main text we have shown a  diagram of a donor/dot array in which the exchange interaction is controlled by a back gate below the donors.  However, a fully planar arrangement in which the donors are closer to the Si/SiO$_{2}$ interface and couple laterally to the dots, as suggested by Carroll and coworkers~\cite{carrolltalk} may be preferable from a fabrication viewpoint. This approach would be closest to current practice for classical silicon circuits, and it would eliminate the need for complex device layers contacted from both sides.  However, for operations such as a logical Hadamard, it would be advantageous for the donor electrons to be able to be moved across more than one site.  Measuring the donors if they are in the same layer as the dots could also lead to a complex routing arrangement for moving electrons to the neighboring donors.   It should be noted that multi-layer heterogeneous integration with micron-scale device registration, as would be necessary for the backgate approach, has been demonstrated~\cite{Li2013}.  Details of layout and fabrication complexity will determine this choice, and are beyond the scope of the present discussion.

Single-qubit operations on the surface code involve applying gates to subsets of the data qubits.  In this architecture the appropriate subset is chosen by selectively swapping to the donors and performing the qubit rotations on them. It would be possible for the measurement donors to do double-duty for the single spin gates, but it may be advantageous to associate a donor with every quantum dot, rather than just the measurement sites, to aid in these operations.  Also, we have discussed measuring and reinitializing a subset of the donors at each step, though measuring all of the donors each time and simply ignoring the unnecessary results would probably be preferable to minimize decoherence.

A further consideration is whether to use the dot electrons as data qubits and the donors for measurement, or the other way around.  There is no particular advantage in terms of the quantum gates, since global single qubit operations can transform the operations appropriately. The major distinction is whether the spin of an electron in a dot or bound to a donor is measured. 

Fast accurate single spin measurement and initialization are required for surface codes, as for other methods of quantum error correction. Spin to charge conversion for spin measurement has been ubiquitous in quantum dot qubit experiments for over a decade~\cite{Elzerman2004}, while direct spin-dependent tunneling for single donor spin readout is a more recent development~\cite{morello10}.  Spin-selective optical excitation of donors may relax requirements on electron temperature~\cite{cheuknew} and placement precision, and optical readout of single donors has been shown~\cite{Yin2013}.  However, the spin readout method is also connected to the arrangement of the donor/dot array.  If the donors are the measurement qubits, all of the donor gates and readout devices (single-electron transistors, SETs~\cite{PhysRevLett.59.109} or quantum point contacts~\cite{PhysRevLett.70.1311} to sense single charges) can be integrated onto the back side of the array.  The transfer gates on the top of the array have a simple structure with this approach.  However, if spin readout is through the dots, then the quantum point contacts (or SETs) for sensing the charge would typically be placed on the surface next to the dots, and the transport of the dot electrons to the nearest neighbor donors becomes more difficult.

%\bibliography{referencesGP}{}

\end{document}